\documentclass[useAMS,usenatbib]{mn2e} 

\usepackage{graphicx} 
\usepackage{mathptmx} 
\def\hst{{\sl HST}}  
\def\teff{$T_{\rm eff}$}  
\def\logg{$\log{g}$}  
\title[Transforming    isochrones    into    the   ACS/WFC    Vega-mag 
system]{Transforming  observational  data  and theoretical  isochrones 
into  the   ACS/WFC  Vega-mag  system\footnotemark[0]\thanks{Based  on 
observations with the NASA/ESA  {\it Hubble Space Telescope}, obtained 
at the Space  Telescope Science Institute, which is  operated by AURA, 
Inc., under NASA contract NAS 5-26555.}} 
\author[L.\ R.\ Bedin et al.]{Luigi R.\ Bedin$^{1}$\thanks{ 
  E-mail:  
  bedin-piotto-momany@pd.astro.it,  
  cassisi-pietrinferni@te.astro.it,  
  castelli@ts.astro.it,  
  jay@eeyore.rice.edu, and 
  ms@astro.livjm.ac.uk.  
}\thanks{ 
current       address:\      European       Southern      Observatory, 
Karl-Schwarzschild-Str.\  2,  85748  b.\  M\"unchen,  Germany,  email: 
lbedin@eso.org}, 
Santi Cassisi$^{2}$,  
Fiorella Castelli$^{3,4}$,  
Giampaolo Piotto$^{1}$,  
Jay Anderson$^{5}$,  
\newauthor  
Maurizio Salaris$^{6}$,  
Yazan Momany$^{1}$, and   
Adriano Pietrinferni$^{2}$. \\ 
$^{1}$Dipartimento  di  Astronomia,  Universit\`a  di  Padova,   
vicolo dell'Osservatorio 2, I-35122 Padova, Italy\\ 
$^{2}$INAF-Osservatorio Astronomico di Collurania,  
via M. Maggini, 64100 Teramo, Italy\\ 
$^{3}$Istituto di Astrofisica Spaziale e Fisica Cosmica, CNR,  
via del Fosso del Cavaliere, I-00133 Roma, Italia\\ 
$^{4}$INAF-Osservatorio Astronomico di Trieste,  
via Tiepolo 11, 34131 Trieste, Italia\\ 
$^{5}$Department of Physics and Astronomy, Mail Stop 108, Rice University,  
6100 Main Street, Houston, TX, 77005, USA\\ 
$^{6}$Astrophysics Research Institute, Liverpool John Moores University,  
12 Quays House, Birkenhead, CH41 1LD, UK. 
} 
 
\begin{document} 
\date{Accepted 2004 December 14. Received 2004 May 20; in original form 2004 May 20} 
\pagerange{\pageref{firstpage}--\pageref{lastpage}} \pubyear{2004} 
\maketitle 
 
\label{firstpage} 
 
\begin{abstract} 
We propose a  zero-point photometric calibration of the  data from the
ACS/WFC on  board the Hubble Space  Telescope, based on  a spectrum of
Vega  and the most  up to  date in-flight  transmission curves  of the
camera.  This calibration is accurate at the level of a few hundredths
of a magnitude.   The main purpose of this effort  is to transform the
entire set of evolutionary models  by Pietrinferni et al.\ (2004) into
a simple  observational photometric system for ACS/WFC  data, and make
them  available to the  astronomical community.   We provide  the zero
points for  the most  used ACS/WFC bands,  and give basic  recipes for
calibrating both the observed data and the models.
 
We also present the Colour Magnitude  Diagram (CMD) from ACS data of 5
Galactic   globular   clusters,   spanning   the   metallicity   range
$-2.2<$[Fe/H]$<-0.04$, and provide  fiducial points representing their
sequences from several  magnitudes below the turnoff to  the red giant
branch tip. The observed sequences are compared with the models in the
newly defined photometric system.
\end{abstract} 
%
\section{INTRODUCTION}  
%
 
After  two and  a half  years of  operation, the  Advanced  Camera for
Surveys (ACS) on board HST has collected a huge number of observations
of  stellar  systems,  and  produced  a  growing  database  of  colour
magnitude diagrams  (CMDs) and luminosity  functions.  In order  to be
able to  exploit fully the  information contained in these  data sets,
and to  extract a number  of fundamental parameters we  are interested
in, such  as metal content,  ages, distances, mass functions,  etc, we
need to  compare the observations with theoretical  models.  The basic
requirement for  this comparison is  the conversion of  stellar models
into the ACS photometric observational domain.  This conversion is the
purpose  of the  present paper.   We  have used  the most  up to  date
information on  the ACS specifications to define  a natural, in-flight
photometric system  and to transform  the Pietrinferni et  al.\ (2004)
models into the same system.
While  it is  foreseeable that  the ACS  photometric  calibration will
improve  somewhat  as  more  standard  fields are  studied  and  cross
compared,  the  in-flight   transmission  efficiency  curves  are  now
reasonably well  established and  we would not  expect them  to change
very  much.  It  therefore  makes  sense to  provide  a conversion  of
theoretical stellar models into  the observational domain so that more
quantitative   and  qualitative   interpretations  can   be   made  of
observations.
 
It   is   worth   mentioning   that  most   astronomical   photometric
investigations  are  based on  some  ``standard'' photometric  system.
Here, by standard,  we mean a photometric system  that has been widely
used  for   a  long  time   in  different  observatories   (e.g.,  the
Johnson-Kron-Cousins, Str\"omgren, Thuan-Gunn, etc systems, see, e.g.,
Landolt-B\"ornstein~1982),  and  possibly   defined  by  a  number  of
standard  stars well  distributed  in  the sky,  whose  flux has  been
carefully measured.
The  calibration to a  standard system  is sometimes  the only  way to
interpret properly the  collected data, e.g.\ when we  need to compare
our  photometric  sequences   with  others  collected  with  different
instruments  or at  different epochs,  or with  standard  results from
theoretical models.   Still, it must  be clearly stated that  when the
transmission curves of the  equipment used to collect the observations
are rather different  from those of any existing  standard system, the
transformation of the data to  a standard system is difficult, and can
be unreliable, particularly for  extreme stars (i.e., extreme colours,
unusual spectral type, high reddening, etc).
 
A  good example  of a  problem that  one can  encounter in  using even
slightly  different filter  sets at  the  same telescope  is given  in
Momany et al.\ (2003).
 
Many  of the ACS  filters differ  substantially from  the ``standard''
filters, and  the high photometric  precision possible with  ACS makes
the systematic  differences more significant.   Thankfully, the filter
and instrumental  response are better characterized for  ACS than they
are for the  typical ground-based observatory, so that  it is possible
to  avoid standard  systems altogether  and compare  the data  and the
model  in the  same system:\  the observational  domain.  All  that is
required  for this  is to  determine an  empirical zeropoint  for each
filter  by  observing  a  reference  star  with  a  well-characterized
spectrum  in  order  to  characterize accurately  the  throughput  and
response of the detector for  each filter.  The theoretical models can
then  be  integrated over  the  same  spectral  response and  compared
directly with  the data.  Such  a procedure for comparing  models with
data is  far preferable  to the typical  procedure of  converting both
models and data into a ``standard'' system.
 
Therefore,  we calculated  a set  of empirical  zero points  that will
allow direct comparison of  observational data and theoretical models.
We have based  our calibration on the reference  spectrum of Vega, and
so  we will  refer to  our photometric  system as  the  ``ACS Vega-mag
system''.  In  our derivation  of zero points,  we used  the in-flight
sensitivity, measured  by Sirianni et  al.\ (2002, and  Sirianni 2003,
private communication).
 
In this paper we have focused our attention on  the Wide Field Channel 
(WFC)  of the ACS, but  the same procedure  can be used to measure the 
photometric zero points for the High Resolution Channel (HRC). 
 
In  the  next Section,  we  describe  briefly  the total  transmission
curves, the procedure followed to obtain the zero points, and give the
aperture corrections.   In Section 3, we briefly  describe the adopted
stellar evolution  models, our procedure to transform  the models into
the  observational plane,  and the  method used  to take  into account
interstellar absorption.  In Section 4 we give the zero points and the
formul\ae ~  to convert both  the theoretical tracks and  the observed
data to the ACS Vega-mag  photometric system.  Section 5 describes the
HST archive  data used to test  the models, and shows,  as an example,
how these data have  been photometrically calibrated and compared with
our models in the ACS Vega-mag plane.  Conclusions follow in Section 6
together with instructions on how to download the transformed models.
 
 
%
\section{DEFINING THE ACS VEGA-MAG PHOTOMETRIC SYSTEM}  
%
 
In this section, we define our photometric system and will refer to it
as the ACS(/WFC) Vega-mag system.
 
%
\subsection{Total transmission of the \hst$+$WFC/ACS}  
%
 
The first step  toward the definition of our  photometric system is to 
calculate  the  total   transmission  curves,  T($\lambda$)  for  each 
individual   band.   These   are  the   products  of   the  individual 
transmission curves, t$_i(\lambda)$, for 5 different terms. 
$$ T(\lambda) = t_1(\lambda) \times ... \times t_5(\lambda)$$  
In detail, the t$_i(\lambda)$ are: $t_1$ the transmission curve of the 
window  of the  dewar, $t_2$  the transmission  curve of  the filters, 
$t_3$ the  reflectivity of the  WFC/ACS mirrors (3 mirrors),  $t_4$ an 
average of the two CCD quantum efficiencies\footnote{ 
These  curves  are  available  at the   STSCI web-site   {\sf  http:// 
www.stsci.edu     /hst   /acs  /analysis            /reference\_files/ 
wfc\_synphottable\_list.html}. 
} and $t_5$ the transmission curve  of the Optical Telescope 
Assembly (OTA) of HST\footnote{ 
This  last component has been obtained  from  the web-site {\sf ftp:// 
ftp.stsci.edu /cdbs /cdbs2 /comp /ota/  } tabulated  in the file  {\sf 
hst\_ota\_007\_syn.fits}. 
}.  
In  order to outline  our  procedure, we   consider two  commonly used 
filters, F606W and F814W. We will extend the  results to other filters 
in Section 4. 
All the individual transmission curves are shown in Fig.\ 1, where two 
thick lines show the total system transmissions for the two considered 
filters. 
 
The   transmission   curves   measured   in  the   laboratory   differ
significantly from  those found on-orbit  (Sirianni et al.\  2002). In
the  following we  will use  {\em  only} the  on-orbit adjusted  total
transmission   curves   obtained   by   Sirianni  et   al.\   (private
communication, paper in preparation 2004).
The  uncertainities for  these  total transmission  curves are  within
0.5\%  (De Marchi  et  al.\ 2004\footnote{ISR  ACS 2004-08,  published
  after  the submission  of the  present work.},  Sirianni et  al.\ in
preparation 2004).
 
%
%
\begin{figure} 
\includegraphics[width=9cm]{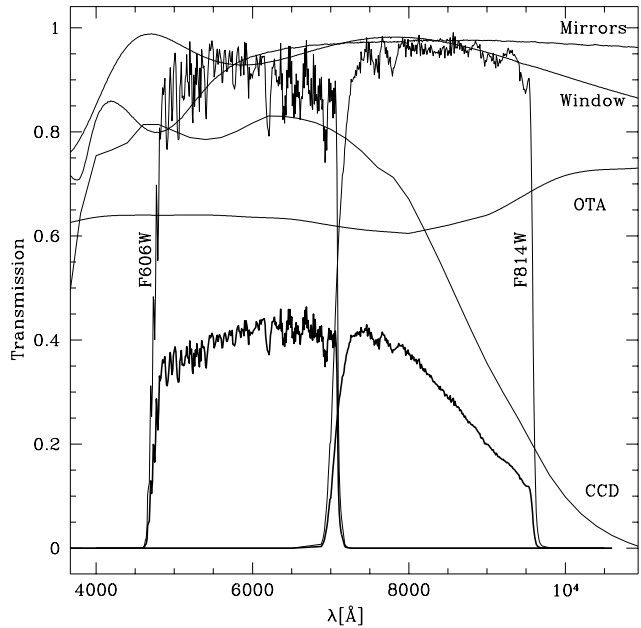} 
\caption{ACS/WFC transmission  curves for the window of  the dewar, of
the WFC/ACS's Mirrors,  of the CCD, of the  Optical Telescope Assembly
(OTA),  and  finally  the  transmission  curves for  the  two  ACS/WFC
filters: F606W  and F814W. The resulting total  transmission curve for
the two filters is shown as a thick line. }
\end{figure} 
 
%
\subsection{The Determination of Zero Points}  

We describe below the procedure used to define the  zero points of the 
system. 
 
We  have used  the spectrum  of Vega  from {\sf  ftp:// ftp.stsci.edu/
cdbs/   cdbs2/  grid/  k93models/   standards/  vega\_reference.fits}.
Details of  this spectrum can  be found at  {\sf http://www.stsci.edu/
instruments/  observatory/ PDF/  scs8.rev.pdf}; hereafter  it  will be
denoted  as  $F^0(\lambda)$,  in   units  ${\rm  erg  ~s^{-1}  cm^{-2}
\AA^{-1}}$.
  
The  integrated flux of  the assumed spectrum  of Vega under the total 
transmission curve $T$ of the filter F606W $+$ instrument is: 

$$   I^0({\rm F606W}) = \int^{\infty}_{-\infty} ...   
\simeq \int^{\lambda_{\rm max}}_{\lambda_{\rm min}}   
F^0 (\lambda) T_{\rm F606W}(\lambda)  d\lambda = $$ 
$$ = 2.7591\times10^{-6} ~ 
{\rm erg ~ s^{-1} cm^{-2}}. $$ 
This number  is the quantity that theoreticians  will need  to rescale 
their models to the photometric system we adopt (see Sec.\ 4). 
 
The estimated number of expected photo-electrons ($\rm e^-$) is: 
$$ N^0_{\rm e^-}({\rm F606W}) = \int^{\infty}_{-\infty} ...  
\simeq \int^{\lambda_{\rm max}}_{\lambda_{\rm min}}   
\frac{F^0 (\lambda)}{e_\gamma (\lambda)} T_{\rm F606W}(\lambda)   
d\lambda  = $$ 
$$ = 795,861 ~ {\rm e^- ~ s^{-1} cm^{-2},}$$  
where the photon energy $e_\gamma (\lambda)$ is given by  
$$e_\gamma (\lambda) = hc / \lambda ~~ {\rm erg}.$$  
 
Therefore, for  Vega, the  counts detected  by  the system  in digital 
numbers (DN) would be: 
$$I_{\rm DN}^0({\rm F606W}) = N^0_{\rm e^-}({\rm F606W}) \times A_{\rm 
eff}  / {\rm GAIN}  = $$ 
$$ = 3.6004 ~ \times10^{10} / {\rm GAIN ~~~~ DN ~ s^{-1},}$$ 
where $A_{\rm eff}$ is the effective  collecting area of the telescope 
(45,239 cm$^2$, with the secondary obscuration already accounted for), 
and GAIN is the conversion factor from e$^-$ to DNs. 
 
At this point, we can define the zero point of our magnitude system in 
the F606W band by imposing $m_{\rm F606W}\equiv0$ for the magnitude of 
Vega (with the assumed spectrum): 
$$   Zp^{\rm   F606W}  \equiv 2.5\,\log_{10}\frac{  I_{\rm  DN}^0({\rm  
F606W})}{[{\rm DN ~ s}^{-1} ]} = 26.391 $$  
For simplicity, we  have defined this zero point  for GAIN $=$  1 $\rm 
e^-/DN$. 
%
 
Since   {\sf  FLT}   images   provided  by   STSCI   are  already   in
photo-electrons (e$^-$) rather than in DNs (note that {\sf DRZ} images
are given  in e$^-$  s$^{-1}$) hereafter, to  avoid any  confusion, we
will not consider the GAIN, and  we will refer to $I_{\rm e^-}$ rather
than to $I_{\rm DN}$.
 
In an analogous way  we can define the  zero point  of the 
F814W band: 
$$  Zp^{\rm F814W} = 25.492, $$  
and in a similar way for all the other filters.   
%
 
 
%
\subsection{Calibration of the Observations}  
%
 
The formula that observers should  use to calibrate their observations 
into our WFC/ACS Vega-mag system magnitudes in the F606W band is: 
$$ m_{\rm F606W} \equiv -2.5 ~ \log_{10}\ \frac{I_{\rm e^-}}{\rm  
exptime} + Zp^{\rm F606W} - Z_{\rm Aperture} $$  
where  exptime is  the exposure  time, and  $Z_{\rm Aperture}$  is the
aperture correction.   In the  next section we  discuss in  detail the
aperture correction, which we will always treat as a positive quantity
to  be subtracted.  This  calibration is  intended for  the officially
calibrated  {\sf DRZ}  images which  are already  corrected  for: flat
field, dark, bias, and geometrical distortion.\footnote{
The images are  still affected, at least by  CTE, and charge diffusion
(see Riess \& Mack, ISR 2004-06, and Krist, ISR 2003-06).
In Section 5.1 we use {\sf FLT} images, and describe how we linked our 
photometry  for those  images to  the officially  corrected  {\sf DRZ} 
images. 
}  
 
%
\subsection {Aperture Corrections}  
%
 
Breathing  of the  telescope  tube, the  complex  optical system,  and
charge diffusion induce temporal and  spatial variation of the core of
the point-spread function (PSF), such that the $1^{\rm st}$-ring ratio
of the diffraction figure changes  considerably across the chip and at
a minor level with orbital phase.
 
For these  reasons, we choose to refer  our zero points to an infinite 
aperture.  That makes it easier for  those studying surface photometry 
of extended sources to calibrate their magnitudes to this system. 
The aperture corrections for extended sources are null, being the Zero 
Points of the system referred to an infinite aperture:\footnote{ 
%
Note, that this is really only  true for infinite extended sources, or 
sources larger than the widest part of the PSF halo. } 
$$ Z_{\rm Aperture} = 0. $$  
On the other  hand, in practice all  the energy of  a point source  is 
contained  within $4''$   ($\sim$  80 pixels),  and  $\sim93\%$ within 
$0.''5$ ($\sim$ 10   pixels).   
In dense environments, where  crowding makes an aperture correction to 
an  infinite aperture  impossible, one  has to  rely on  the in-flight 
encircled energy.\footnote{ 
This is  tabulated at the  STSCI web-site: {\sf  http:// www.stsci.edu 
/hst/acs/analysis/ reference\_files/ wfc\_synphottable\_list.html}} 
This means  that one  has to correct  the instrumental  magnitude (for 
example  the PSF-fitted  magnitude) to  the magnitude  at  a specified 
aperture $r$  by adding a  $\Delta m_{\rm PSF-AP({\it r})}$,  and then 
sum  an additional quantity  $\Delta m_{\rm  AP({\it r})-AP(\infty)}$, 
obtained  from the  tabulated encircled  energy in  order to  link the 
magnitudes to the  infinite aperture for which we  have calculated the 
Zero Points. 
 
%
%
In summary,   
$$ Z_{\rm Aperture} =  
\Delta m_{\rm PSF-AP({\it r})} +  
\Delta m_{\rm AP({\it r})-AP(\infty)}.$$  
 
In  Table~1  we  give  the $\Delta  m_{\rm  AP({\it  r})-AP(\infty)}$ 
corrections to be added to  the zero points for two circular apertures 
of radius $0.''3$ ($\sim$6 pixels) and $0.''5$ (as in Holtzman et al.\ 
1995) and for the  different filters.  For clarity, we  also report in 
Tab.\ 1 the correction to  zero points for an infinite aperture, which 
is zero by definition. 
 
%
%
\begin{table} 
\begin{center}  
\begin{tabular}{c|ccc}  
$ FILTER $     & $\Delta m_{\rm AP({\infty})-AP(\infty)}$   
                         & $\Delta m_{\rm AP(0''.5)-AP(\infty)}$   
                                   & $\Delta m_{\rm AP(0''.3)-AP(\infty)}$  \\  
\hline  
 F435W                   &     0.000      &  0.095        & 0.132        \\  
 F475W                   &     0.000      &  0.084        & 0.120        \\  
 F502N                   &     0.000      &  0.078        & 0.114        \\  
 F550M                   &     0.000      &  0.076        & 0.110        \\  
 F555W                   &     0.000      &  0.076        & 0.110        \\  
 F606W                   &     0.000      &  0.079        & 0.115        \\  
 F625W                   &     0.000      &  0.080        & 0.118        \\  
 F658N                   &     0.000      &  0.079        & 0.123        \\  
 F660N                   &     0.000      &  0.079        & 0.123        \\  
 F775W                   &     0.000      &  0.078        & 0.123        \\  
 F814W                   &     0.000      &  0.079        & 0.130        \\  
\end{tabular}  
\end{center}  
\caption{$\Delta  m_{\rm AP({\it  r})-AP(\infty)}$  corrections to  be 
added to the Zero Points. } 
\label{apes}  
\end{table}  
 
For  point sources,  the $\Delta  m_{\rm PSF-AP({\it  r})}$ correction
should  be estimated in  each individual  image, remembering  that the
smaller the aperture ($r$), the more the photometry is affected by the
problems  mentioned at the  beginning of  this section  (i.e.\ spatial
variability of  the PSF  and charge diffusion).   See Section  5.2 for
uncertainties on the values of $\Delta m_{\rm PSF-AP({\it r})}$.
 
Moreover, there appears  to be a second-order effect  which depends on 
the  colour of the  stars and  on the  filter used  (Sirianni, private 
communication,  paper  in  preparation).   In  practice,  the  present 
calibration is  useful only if the  degree of accuracy  desired by the 
observers is no  better than 2-3 hundredths of  a magnitude.  However, 
taking into account all the problems  of ACS, it seems hard to achieve 
a much higher accuracy. 
It  should  be noted  that  the  photometric  stability of  WFPC2  was 
certainly  not any  better  than  this, but  the  higher precision  of 
WFC/ACS  makes systematic  errors (on  the zero  points)  the dominant 
source of error in the final photometry. 
 
We have decided  not to deal with the two  filters:\ F850LP and F892N, 
because they are much more problematic (see Sirianni, in preparation). 
 
 
%
\section{TRANSFORMING THE STELLAR MODELS INTO THE ACS VEGA-MAG SYSTEM}  
%
 
%
\subsection{The Stellar Evolution Models}  
%
 
In  this  work  we  adopt  the  stellar-evolution  model  database  in
Pietrinferni et al.\ (2004); for a detailed description of the models,
as well as  a comparison with other stellar  model libraries available
in the literature,  we refer the interested reader  to this reference.
For  the   purposes  of   this  paper,  we   briefly  list   the  main
characteristics of this model library.
 
Evolutionary tracks are available for up  to 41 mass values at each of
10 selected initial chemical compositions. The minimum mass considered
is roughly $0.5  M_\odot$, while the maximum value  is always equal to
$10M_\odot$.  The whole set of evolutionary computations has been used
to compute  isochrones for  a large age  range, from $\sim  40$~Myr to
$\sim 14$~Gyr, covering all stellar evolution phases from the Zero Age
Main Sequence  up to the first  thermal pulse on  the Asymptotic Giant
Branch or  to C  ignition. For each  chemical composition, a  Zero Age
Horizontal  Branch (ZAHB)  locus,  as well  as post-ZAHB  evolutionary
tracks were also calculated for  a large range of masses, by employing
the He-core mass  and chemical profiles in the  H-rich envelope of the
Red Giant Branch (RGB) progenitor having an age at the He-flash of the
order of $\sim12-13$~Gyr.
 
The set  of models  has been computed  for metallicities in  the range
$0.0001\le  {Z} \le  0.04$  for  both a  scaled-solar  mixture and  an
$\alpha$-enhanced   mixture   ($<$[$\alpha$/Fe]$>$$=$0.4;   paper   in
preparation).
 
The  adopted  reference scaled-solar  heavy  element  mixture is  from
Grevesse \& Noels~(1993).  As  for the He-abundance, the models employ
a primordial mass fraction of  $Y_0$=0.245 , based on the estimates by
Cassisi et  al.\ (2003, see also  Salaris et al.\  2004) obtained from
the  $R$-parameter  method  applied  to  a large  sample  of  Galactic
globular clusters (GGCs).  This $Y_0$  value is in fair agreement with
recent independent  determinations of the  cosmological baryon density
from  the WMAP  results  (Spergel  et al.\  2003).   To reproduce  the
initial solar  He-abundance obtained from  an appropriately calibrated
solar  model,  an  He  enrichment  ratio  $dY/dZ=1.4$  has  been  used
(Pietrinferni et al.\ 2004).

All models have been  computed with outer boundary conditions obtained 
by  integrating the atmospheric  layers with  the Krishna-Swamy~(1966) 
$T(\tau)$   relationship.    Superadiabatic   convection  is   treated 
according to the Cox \& Giuli~(1968) formalism of mixing-length theory 
(B\"ohm-Vitense~1958), and the  mixing-length parameter has been fixed 
by the solar calibration ($ml=1.713$) and kept constant for all masses 
during all evolutionary phases; mass loss is included by employing the 
Reimers formula  (Reimers~1975) with the free parameter  $\eta$ set to 
0.4. 
 
At  each chemical  composition,  models have  been  computed with  and 
without  overshooting  from  the  convective  cores  during  the  main 
sequence      (MS)      phase.       In     the      former      case, 
$\lambda_{OV}=0.20\times{H_P}$  for masses  larger  than 1.7$M_\odot$, 
where $\lambda_{OV}$  is the extension  of the overshooting  region in 
units of the local pressure scale height.  For stars less massive than 
$1.1  M_\odot$,  $\lambda_{OV}=0$,  while  in the  intermediate  range 
$\lambda_{OV}$      varies     according      to      the     relation 
$\lambda_{OV}$$=$$(M/M_{\odot}-0.9)/4$.   This choice allows  a smooth 
variation  of the  isochrone turn-off  (TO) morphology,  and  a smooth 
decrease to  zero of the  convective core mass.   This parametrization 
also provides a good match to the TO morphology in the CMD of Galactic 
open   clusters   of  different   ages.    Induced  overshooting   and 
semi-convection during the central  He-burning phase are accounted for 
following Castellani et al.\ (1985). 
 
Radiative opacities from the OPAL tables (Iglesias \& Rogers~1996) for 
temperatures larger   than     $10^4$  K, and    from    Alexander  \& 
Ferguson~(1994)  for      lower temperatures  have     been  employed, 
supplemented   by opacities for   electron degenerate  matter computed 
following Potekhin~(1999).   The    relevant energy   loss  rate  from 
plasma-neutrino processes comes from  Haft,  Raffelt \&  Weiss~(1994), 
whereas for all other  processes the reader is  referred to Cassisi \& 
Salaris~(1997); the nuclear reaction rates are from the NACRE database 
(Angulo  et    al.\      1999),   with   the      exception  of    the 
$^{12}$C$(\alpha,\gamma)^{16}$O   reaction,  in   which  case the more 
accurate recent determination by Kunz et al.\ (2002) has been adopted. 
Electron screening is treated according to Graboske et al.\ (1973). 
   
The accurate Equation  of State (EOS) by A.\ Irwin  has been used.  An 
exhaustive description  of this EOS  is in preparation (Irwin  et al.\ 
2004) but a brief discussion of  its main characteristics can be found 
in Cassisi et al.\ (2003).  This EOS covers the full stellar structure 
for the whole  evolutionary phase and mass range  spanned by the model 
database;  in Pietrinferni et  al.\ (2004)  and Bahcall,  Serenelli \& 
Pinsonneault (2004) it is shown that this EOS and the latest version of 
the OPAL  EOS (Rogers  2001) provide almost  identical results  in the 
case of the Sun. 
%
%
The accuracy and reliability  of the present evolutionary scenario has
been successfully  tested in  Cassisi et al.\  (2003), Riello  et al.\
(2003) and Salaris et  al.\ (2004) in the regime  of Galactic globular
clusters, and  by Pietrinferni et al.\  (2004) in the  regime of field
stars and open clusters.
%
%
%
Cassisi  et  al.~(2003)  and  Pietrinferni  et  al.~(2004)  have  also
compared our adopted set of theoretical isochrones with other existing
isochrone  databases widely  cited  in the  current literature.   This
comparison  provides  an  estimate  of the  present  uncertainties  in
stellar  evolution models  due to  different choices  for  the stellar
input physics (see also  Chaboyer~1995).  The main conclusions of this
exercise are summarized below.
 
Isochrones for typical globular cluster  ages (of the order of 10~Gyr)
show a remarkable consistency for  the prediction of the Main Sequence
location  and  Turn  Off   luminosity  at  a  given  initial  chemical
composition, whereas  the Red  Giant Branch effective  temperature and
Zero Age  Horizontal Branch brightness  display values that  differ by
amounts of  $\sim$100-200~K and $\sim$0.1~mag,  respectively. Globular
cluster  ages  estimated from  the  difference  between  Turn Off  and
Horizontal  Branch  brightness  would  consequently  be  uncertain  by
$\sim$10 $\%$ (Chaboyer~1995 determined  a similar uncertainty, of the
order of 15\%).
 
Isochrones  for ages  lower than  $\sim$4-5~Gyr show  again consistent
Main  Sequence   locations,  but  differing   Turn  Off  luminosities,
effective temperatures, and Helium burning luminosities, due mainly to
the  uncertain treatment of  the overshooting  from the  Main Sequence
stars' convective  cores. These uncertainties cause  variations in the
estimated age of stellar populations within this age range, of at most
$\sim$30  \%.    Also  in  this   age  range  red   giants'  effective
temperatures show a spread of $\sim$100-200~K.

\subsection{The Transformation to the Observational Plane}  

Here  we describe  the procedure  we  have followed  to transform  our 
models into the ACS Vega-mag system defined in Sect.~2. 
 
We computed ACS colour indices  by using the homogeneous set of ODFNEW
model  atmospheres  and  synthetic fluxes\footnote{Available  at  {\sf
http:// wwwuser.oat.ts.astro.it/ castelli/ grids.html}.}  (Castelli \&
Kurucz 2003) computed with the  Kurucz ATLAS9 code. We generated grids
of indices for the following  values of [Fe/H]: 0.5, 0.2, 0.0, $-$0.5,
$-$1.0, $-$1.5, $-$2.0, $-$2.5.
 
In each grid point, \teff\  ranges from 3500 to 50,000\,K, \logg\ from
0.0     to    5.0     and    the     microturbulent     velocity    is
$\xi$$=$2.0\,km\,s$^{-1}$.   We tabulated for  each \teff\  and \logg\
gridpoint the  BC$_{V}$ bolometric correction, the  $V$ magnitude, and
the  colour indices  $V-m_{\rm F435W}$,  $V-m_{\rm  F475W}$, $V-m_{\rm
F555W}$, $V-m_{\rm F606W}$,  $V-m_{\rm F625W}$, $V-m_{\rm F775W}$, and
$V-m_{\rm  F814W}$.  The  ACS magnitudes  were computed  by  using the
WFC/ACS transmission curves described  in Sect.\ 2.1, while we adopted
the $V$ passband from Bessel (1990).

We adopted $V  = 0.03$ for Vega and set  all colour indices ($V-m_{\rm 
filter}^{\rm ACS}$) to zero, so  that the Vega ACS magnitudes would be 
equal to 0.00 in all passbands. 
The Vega model and flux are  the ATLAS9 model and the ATLAS9 flux from 
Castelli   \&   Kurucz   (1994)   with   parameters   \teff$=$9550\,K, 
\logg$=$3.95, [M/H]$=$$-$0.5, $\xi$$=$2\,km\,s$^{-1}$. 
Note that this  model has  been built  specifically  to reproduce  the 
observations of the spectrum used in Sect.\ 2.2. 
The absolute zero-point for the  $V$ magnitude was fixed by adding the 
constant $-$21.10 mag to the  computed $V$ for Vega (Bessell, Castelli 
\& Plez 1998). 

In  Table~2   we  show  an  example  of   the  theoretical  isochrones
transformed into  the ACS Vega-mag photometric system.   The header of
the isochrone table provides information  about the number of lines in
the file ($N_P$), the total  metallicity $\rm [M/H]$, the abundance by
mass of  metals ($Z$) and helium  ($Y$), and the  corresponding age in
billions  of years,  $t(Gyr)$.   The  content of  the  columns in  the
following lines is the following  (from left to right): 1) the initial
mass of each star, 2) the  current mass at the specified age (computed
accounting for mass loss -  see previous section), 3) the logarithm of
the  surface  luminosity in  solar  units,  4)  the logarithm  of  the
effective temperature,  5) the ACS  $M_{\rm F435W}$ magnitude,  6) the
ACS $M_{\rm  F475W}$ magnitude, 7) the ACS  $M_{\rm F555W}$ magnitude,
8) the  ACS $M_{\rm  F606W}$  magnitude, 9)  the  ACS $M_{\rm  F625W}$
magnitude, 10) the ACS $M_{\rm  F775W}$ magnitude, 11) the ACS $M_{\rm
F814W}$ magnitude.
 
The  entire set of  transformed isochrones,  as well  as the  ZAHB and 
post-ZAHB  evolutionary tracks  can  be retrieved  from the  following 
URL:\ {\sf http:// www.te.astro.it/BASTI/index.php}. 
 
%
%
\begin{table*}  
\begin{center}  
{\small 
\begin{tabular}{c}  
 Isochrone by Pietrinferni - Cassisi - Salaris - Castelli 2004     Standard Model           \\ 
 Scaled solar model \& transformations for ACS           (Castelli 2004)                    \\ 
 $Np\!=\!2000$,~~ $[M/H]\!=\!-0.659$,~~ $Z\!=\!0.0040$,~~ $Y\!=\!0.251$,~~ $t(Gyr)\!=\!11.0000$ \\ 
\end{tabular}  
} 
{\tiny  
\begin{tabular}{ccccccccccc}  
 $\rm (M/M_\odot)_{in}$  
            & $\rm (M/M_\odot)$    
                         & $\rm \log{L/L_\odot}$  
                                    & $\rm \log{T_{eff}}$   
                                              & $M_{\rm F435W}$  
                                                      & $M_{\rm F475W}$  
                                                              & $M_{\rm F555W}$  
                                                                      & $M_{\rm F606W}$  
                                                                              & $M_{\rm F625W}$  
                                                                                      & $M_{\rm F775W}$  
                                                                                              & $M_{\rm F814W}$ \\ 
\hline 
.5000000000 &.4999030025 & -1.20676 & 3.62790 & 9.540 & 9.063 & 8.418 & 8.024 & 7.693 & 7.069 & 6.982 \\ 
.5011654487 &.5011233468 & -1.20259 & 3.62829 & 9.525 & 9.048 & 8.405 & 8.012 & 7.681 & 7.059 & 6.972 \\ 
.5023308974 &.5023308974 & -1.19841 & 3.62868 & 9.510 & 9.034 & 8.392 & 8.000 & 7.670 & 7.049 & 6.962 \\ 
     ...    &      ...   &    ...   &    ...  &   ... &  ...  &  ...  &  ...  &  ...  &  ...  &  ...  \\ 
.9066248329 &.6047402926 &  2.89310 & 3.60793 & -0.175&-0.871 & -1.635&-2.045 & -2.382&-3.053 &-3.153 \\ 
.9066266265 &.6037663031 &  2.92275 & 3.60503 & -0.190&-0.899 & -1.675&-2.093 & -2.435&-3.116 &-3.218 \\ 
.9066284200 &.6026715201 &  2.95336 & 3.60202 & -0.204&-0.926 & -1.715&-2.143 & -2.490&-3.181 &-3.285 \\ 
\end{tabular}  
} 
\end{center}  
\caption{Example   of   an  isochrone   available   at  {\sf   http:// 
www.te.astro.it/ BASTI/ index.php}. } 
\label{iso}  
\end{table*}  
%
%
%
 
%
\subsection{Reddening in the WFC/ACS Vega-mag system}  
 
Following the  recipe and  the motivations given  by Holtzman  et al.\ 
(1995) we  also computed  the interstellar  absorption in  the WFC/ACS 
bands as a  function of $E(B-V)$.  By using  the extinction curve from 
Table~1 in  Mathis (1990)  for  $R_V=3.1$ we  computed for  different 
values  of $E(B-V)$  the interstellar  reddening $A_{\lambda}$  at the 
wavelengths of the ATLAS9  flux, the corresponding reddened fluxes and 
the reddened WFC/ACS magnitudes.   The $A_{\rm filter}$ extinctions in 
the WFC/ACS filters were  obtained as differences between the reddened 
and unreddened $m_{\rm filter}$ magnitudes. 
 
Tables~3  and 4  give the values  of the  $A_{\rm filter}$ for  a cool 
star and a hot star respectively.  We give these values for the main 
filters, F435W, F475W, F555W, F606W,  F625W, F775W, and F814W. We note 
how these values are very close to those tabulated by Holtzman et al.\ 
(1995) for WFPC2's similar filters. 
 
%
%
\begin{table*}  
\begin{center}  
{\tiny 
%
\begin{tabular}{c}  
$T_{\rm eff}=4,000$K, $\log{g}=4.50$, $\rm [M]=0.00$,  $v_{\rm turb}=2.00$,  $l/H=1.25$  \\ 
\begin{tabular}{cccccccc}  
$E(B-V)$& $A_{\rm F435W}$ & $A_{\rm F475W}$ & $A_{\rm F555W}$ & $A_{\rm F606W}$ & 
                       $A_{\rm F625W}$ & $A_{\rm F775W}$ & $A_{\rm F814W}$ \\ 
\hline 
0.00 &  0.000 &  0.000 &  0.000 &  0.000 &  0.000 &  0.000 &  0.000 \\ 
0.05 &  0.202 &  0.182 &  0.158 &  0.140 &  0.132 &  0.099 &  0.092 \\ 
0.10 &  0.404 &  0.365 &  0.316 &  0.279 &  0.263 &  0.198 &  0.184 \\ 
0.15 &  0.606 &  0.546 &  0.474 &  0.419 &  0.394 &  0.297 &  0.276 \\ 
0.20 &  0.808 &  0.728 &  0.632 &  0.558 &  0.526 &  0.396 &  0.367 \\ 
0.25 &  1.010 &  0.909 &  0.790 &  0.696 &  0.657 &  0.495 &  0.459 \\ 
0.30 &  1.212 &  1.090 &  0.947 &  0.834 &  0.788 &  0.594 &  0.550 \\ 
0.35 &  1.413 &  1.271 &  1.104 &  0.972 &  0.919 &  0.692 &  0.641 \\ 
0.40 &  1.614 &  1.451 &  1.262 &  1.110 &  1.050 &  0.791 &  0.732 \\ 
0.45 &  1.815 &  1.631 &  1.419 &  1.248 &  1.180 &  0.890 &  0.823 \\ 
0.50 &  2.017 &  1.811 &  1.575 &  1.385 &  1.311 &  0.988 &  0.913 \\ 
0.55 &  2.217 &  1.990 &  1.732 &  1.522 &  1.442 &  1.086 &  1.004 \\ 
0.60 &  2.418 &  2.169 &  1.889 &  1.658 &  1.572 &  1.185 &  1.094 \\ 
0.65 &  2.619 &  2.348 &  2.045 &  1.794 &  1.702 &  1.283 &  1.184 \\ 
0.70 &  2.819 &  2.527 &  2.201 &  1.930 &  1.833 &  1.381 &  1.274 \\ 
0.75 &  3.020 &  2.705 &  2.357 &  2.066 &  1.963 &  1.479 &  1.364 \\ 
0.80 &  3.220 &  2.883 &  2.513 &  2.202 &  2.093 &  1.577 &  1.454 \\ 
1.00 &  4.020 &  3.592 &  3.136 &  2.741 &  2.612 &  1.968 &  1.811 \\ 
1.25 &  5.016 &  4.472 &  3.911 &  3.409 &  3.260 &  2.454 &  2.254 \\ 
1.50 &  6.010 &  5.347 &  4.682 &  4.072 &  3.905 &  2.939 &  2.693 \\ 
1.75 &  7.001 &  6.215 &  5.450 &  4.730 &  4.547 &  3.422 &  3.128 \\ 
2.00 &  7.990 &  7.077 &  6.216 &  5.382 &  5.188 &  3.902 &  3.559 \\ 
2.25 &  8.975 &  7.935 &  6.978 &  6.031 &  5.827 &  4.381 &  3.987 \\ 
2.50 &  9.959 &  8.788 &  7.737 &  6.675 &  6.463 &  4.858 &  4.411 \\ 
2.75 & 10.940 &  9.636 &  8.494 &  7.315 &  7.098 &  5.333 &  4.832 \\ 
3.00 & 11.919 & 10.480 &  9.249 &  7.951 &  7.731 &  5.806 &  5.250 \\ 
3.25 & 12.896 & 11.320 & 10.001 &  8.585 &  8.362 &  6.277 &  5.664 \\ 
3.50 & 13.871 & 12.157 & 10.751 &  9.215 &  8.991 &  6.747 &  6.075 \\ 
3.75 & 14.844 & 12.991 & 11.499 &  9.843 &  9.619 &  7.215 &  6.483 \\ 
4.00 & 15.815 & 13.822 & 12.244 & 10.468 & 10.245 &  7.681 &  6.888 \\ 
4.25 & 16.785 & 14.650 & 12.988 & 11.091 & 10.870 &  8.145 &  7.291 \\ 
4.50 & 17.753 & 15.476 & 13.730 & 11.711 & 11.493 &  8.608 &  7.690 \\ 
4.75 & 18.719 & 16.299 & 14.470 & 12.329 & 12.115 &  9.070 &  8.088 \\ 
5.00 & 19.684 & 17.120 & 15.209 & 12.946 & 12.735 &  9.530 &  8.483 \\ 
 & & & & & & & \\ 
 & & & & & & & \\ 
 & & & & & & & \\ 
\end{tabular}  
\end{tabular} 
} 
\end{center}  
\caption{Extinctions (in  mag) in different ACS filters  as a function 
of the colour excess $E(B-V)$ of a cool star. } 
\label{cool}  
\end{table*}  
 
%
%
\begin{table*} 
\begin{center}  
{\tiny 
%
\begin{tabular}{c}  
$T_{\rm eff}=40,000$K, $\log{g}=4.50$, $\rm [M]=0.00$,  $v_{\rm turb}=2.00$,  $l/H=1.25$  \\ 
\begin{tabular}{cccccccc}  
$E(B-V)$& $A_{\rm F435W}$ & $A_{\rm F475W}$ & $A_{\rm F555W}$ & $A_{\rm F606W}$ &  
                                $A_{\rm F625W}$ & $A_{\rm F775W}$ & $A_{\rm F814W}$ \\ 
\hline 
0.00 &  0.000 &  0.000 &  0.000 &  0.000 &  0.000 &  0.000 &  0.000 \\ 
0.05 &  0.212 &  0.194 &  0.165 &  0.153 &  0.135 &  0.102 &  0.096 \\ 
0.10 &  0.423 &  0.387 &  0.329 &  0.305 &  0.269 &  0.203 &  0.192 \\ 
0.15 &  0.635 &  0.580 &  0.493 &  0.456 &  0.403 &  0.304 &  0.288 \\ 
0.20 &  0.846 &  0.773 &  0.657 &  0.608 &  0.538 &  0.406 &  0.383 \\ 
0.25 &  1.057 &  0.965 &  0.821 &  0.759 &  0.672 &  0.507 &  0.479 \\ 
0.30 &  1.267 &  1.157 &  0.985 &  0.909 &  0.806 &  0.608 &  0.574 \\ 
0.35 &  1.478 &  1.348 &  1.148 &  1.059 &  0.940 &  0.709 &  0.669 \\ 
0.40 &  1.688 &  1.539 &  1.311 &  1.209 &  1.073 &  0.810 &  0.764 \\ 
0.45 &  1.898 &  1.730 &  1.474 &  1.358 &  1.207 &  0.911 &  0.859 \\ 
0.50 &  2.108 &  1.921 &  1.637 &  1.507 &  1.341 &  1.012 &  0.953 \\ 
0.55 &  2.318 &  2.111 &  1.800 &  1.656 &  1.474 &  1.112 &  1.048 \\ 
0.60 &  2.527 &  2.300 &  1.963 &  1.804 &  1.608 &  1.213 &  1.142 \\ 
0.65 &  2.737 &  2.490 &  2.125 &  1.951 &  1.741 &  1.313 &  1.237 \\ 
0.70 &  2.946 &  2.678 &  2.288 &  2.099 &  1.874 &  1.414 &  1.331 \\ 
0.75 &  3.155 &  2.867 &  2.450 &  2.246 &  2.007 &  1.514 &  1.424 \\ 
0.80 &  3.364 &  3.055 &  2.612 &  2.392 &  2.140 &  1.615 &  1.518 \\ 
1.00 &  4.196 &  3.804 &  3.258 &  2.975 &  2.671 &  2.015 &  1.891 \\ 
1.25 &  5.233 &  4.732 &  4.062 &  3.694 &  3.333 &  2.513 &  2.354 \\ 
1.50 &  6.265 &  5.651 &  4.862 &  4.405 &  3.992 &  3.009 &  2.813 \\ 
1.75 &  7.292 &  6.562 &  5.658 &  5.107 &  4.648 &  3.503 &  3.268 \\ 
2.00 &  8.315 &  7.465 &  6.450 &  5.800 &  5.302 &  3.995 &  3.719 \\ 
2.25 &  9.333 &  8.360 &  7.239 &  6.486 &  5.954 &  4.485 &  4.166 \\ 
2.50 & 10.347 &  9.247 &  8.023 &  7.165 &  6.603 &  4.972 &  4.609 \\ 
2.75 & 11.357 & 10.128 &  8.804 &  7.837 &  7.250 &  5.458 &  5.048 \\ 
3.00 & 12.362 & 11.003 &  9.581 &  8.503 &  7.895 &  5.942 &  5.483 \\ 
3.25 & 13.364 & 11.872 & 10.355 &  9.163 &  8.538 &  6.423 &  5.915 \\ 
3.50 & 14.362 & 12.735 & 11.126 &  9.818 &  9.178 &  6.902 &  6.343 \\ 
3.75 & 15.357 & 13.594 & 11.894 & 10.468 &  9.817 &  7.380 &  6.767 \\ 
4.00 & 16.348 & 14.447 & 12.658 & 11.114 & 10.453 &  7.856 &  7.187 \\ 
4.25 & 17.336 & 15.297 & 13.420 & 11.755 & 11.088 &  8.329 &  7.604 \\ 
4.50 & 18.321 & 16.142 & 14.179 & 12.393 & 11.721 &  8.801 &  8.018 \\ 
4.75 & 19.303 & 16.985 & 14.935 & 13.028 & 12.352 &  9.271 &  8.429 \\ 
5.00 & 20.283 & 17.823 & 15.689 & 13.659 & 12.981 &  9.740 &  8.837 \\ 
\end{tabular}  
\end{tabular} 
} 
\end{center}  
\caption{As in Table~3, but for a hot star. } 
\label{hot}  
\end{table*}  
%
\section{ACS VEGA-MAG SYSTEM ZERO POINTS}  
%

Table~5 lists the  zero points for the   ACS Vega-mag system  in other 
useful pass-bands; the  zero points for  all the other filters can  be 
easily obtained following the recipes outlined above. 
 
We  note that this  calibration is  linked to  the chosen  spectrum of 
Vega.  These zero points are probably accurate to 2--3 hundredths of a 
magnitude, and  can (and  hopefully will soon)  be updated,  once more 
accurate data on the ACS performance is available. 
 
The  main  sources of  systematic  error  in  the calibration  of  the
observations  are: residuals  in  correction for  CTE, flat  fielding,
aperture corrections, charge diffusion, and the encircled energy which
seems to depend on stellar colour (Sirianni, private communication).
Still,  the  lack  of  any  other official  and  complete  photometric
calibration of the ACS, useful for the comparison with stellar models,
has led us to calculate and publish this calibration.
 
%
%
\begin{table*} 
\begin{center}  
\begin{tabular}{c|cccccc}  
filter & F435W  & F475W   & F502N    & F550M   & F555W  & F606W  \\  
\hline 
$Zp^{\rm filter}$ 
       & 25.785 & 26.168  & 22.338   & 24.861  & 25.718 & 26.391 \\  
$I^0$(filter)    
       & 2.0994 &  2.7611 & 0.075320 & 0.69396 & 1.6028 & 2.7591 \\  
       &        &         &          &         &        &        \\ 
filter & F625W  & F658N   & F660N    & F775W   & F814W   & ----- \\  
\hline 
$Zp^{\rm filter}$  
       & 25.722 & 22.381  & 21.342   & 25.254  & 25.492  & ----- \\  
$I^0$(filter)    
       & 1.3687 & 0.059770& 0.022895 & 0.72740 & 0.87198 & ----- \\  
\end{tabular}  
\end{center}  
\caption{Natural ACS-VEGA System  Zero Points ($Zp^{\rm filter}$), and 
Fluxes $I^0({\rm filter})$ (in 10$^{-6}$ erg s$^{-1}$ cm$^{-2}$). } 
\label{zps}  
\end{table*}  

In  summary, photometric observations   can  be transformed  into  the 
ACS/WFC  Vega-mag  system  defined  in this   paper   by applying  the 
following relation: 
$$m_{\rm filter}  \equiv -2.5 \,\log_{10}\ \frac{I_{\rm  e^-}}{\rm exptime} +  Zp^{\rm filter}  +$$  
\begin{equation}  
~~~~~~~~~~~~~~~~~~~~~~~~~~~~~~~~~~- \Delta m_{\rm PSF-AP({\it r})}^{\rm filter} 
                    - \Delta m_{\rm AP({\it r})-AP(\infty)}^{\rm filter}; 
\label{eq_obs}  
\end{equation}  
Where $r$ can assume the  values: $\infty$, $0''.5$, or $0''.3$.
Note that the last two positive quantities in Eq.~\ref{eq_obs} are subtracted. 
 
Models can be transformed  into the observational  plane of the  same 
system by applying: 
\begin{equation}  
M_{\rm filter} =   
-2.5 \,\log_{10}\ \frac{ I({\rm filter}) }{ I^0({\rm filter})} .  
\end{equation}  
Here the  meaning of $I({\rm  filter})$, and $I^0({\rm  filter})$ are 
clear from the discussions in previous sections.  In Sect.\ 5.2 we give 
a practical example of how to calibrate the observations. 
 
Our Zero points agree well 
with  those  presented  in  the  recent  ISR  by  De  Marchi  et  al.\ 
(2004-08).\footnote{ 
\sf 
http://\ www.stsci.edu/\ hst/\  acs/\ documents/\ isrs/\ isr0408.pdf.} 
The only  differences are due to  a different choice  of spectrum of 
Vega (Sirianni private comunication).  Nevertheless, for broad filters
the Zero points agree within $\sim$0.005 mag. 
 
 
%
\section{COMPARISON OF MODELS WITH OBSERVATIONS}  
%
 
%
\subsection{Observations and Data Reduction}  
%
 
In order to give a practical  example of our procedures and to provide
a useful test  for our models, we used archival  WFC observations of 5
Galactic globular  clusters (GO-9453, PI:\  Brown), which span  a wide
range  in metallicity ($\rm  -2.2 <  [Fe/H] <  -0.04$).  The  data are
summarized in Table~6.  They consist of 1 short, 1 intermediate, and 1
(relatively) long exposure, in both  F606W and F814W filters, for each
cluster.
  
We carried out  the photometry with algorithms based  on the effective
point-spread function  (ePSF) fitting procedure  described by Anderson
\&  King (2000). We  obtained the  ePSF from  a set  of images  from a
different  project using  the  same filters,  since  the dithering  of
GO-9453 images was unsuitable to constrain the ePSF shape (Anderson \&
King, 2000).
 
The resampling present in the  {\sf DRZ} images provided by STSCI does 
not allow  the use of the  ePSF method, limiting the  precision of the 
output  photometry.   We therefore  performed  our fitting  photometry 
using  {\sf  FLT}  images.   These  images,  affected  by  geometrical 
distortion, have been flat-fielded to preserve surface brightness, but 
not flux.   Therefore, we  must multiply each  pixel by  its projected 
area to obtain a proper flux for each star. 
 
We find slight differences  of flux-scaling between our corrected {\sf
FLT} images  and the corresponding {\sf  DRZ} images (on  the order of
1\% or so).  Since it is the {\sf DRZ} images that have been carefully
calibrated via  the STSCI pipe-line, we should  convert any photometry
measured on {\sf FLT}  images into the corresponding infinite-aperture
fluxes in the {\sf DRZ} images.  The photometry has been calibrated as
described in the previous sections.
%
\subsection{An Example: Calibration of the 47Tuc Photometry}  
%
 
%
For the GO-9453  47Tuc data, the zero point of  the long exposures has
been used to  calibrate the intermediate and short  ones in both F606W
(for which the exposure time is  70s) and F814W filters (for which the
exposure  time is  72s, see  Table~3).  We  determined  the zero-point
difference between our  PSF-fitting photometry and aperture photometry
with a $0''.5$ aperture for each filter.
The  differences  are  $\Delta  m_{\rm PSF-AP(0''.5)}^{\rm  F606W}  =$ 
$+0.15(\pm0.015)$,  and $\Delta  m_{\rm PSF-AP(0''.5)}^{\rm  F814W} =$ 
$+0.25(\pm0.015)$.  To these we  need to add the aperture corrections. 
According  to  Table~1:\  $\Delta  m_{\rm  AP(0''.5)-AP(\infty)}^{\rm 
F606W} =  \Delta m_{\rm AP(0''.5)-AP(\infty)}^{\rm  F814W} =$ $+0.079$ 
(equal for the two filters, at this apertures).  In summary, from Eq.\ 
(\ref{eq_obs}): 
%
$$ m_{\rm  F606W} \equiv -2.5  ~ \log_{10}\ \frac{I_{\rm e^-}}{\rm 
  70} + 26.391 - 0.15 - 0.079 $$ 
$$ m_{\rm  F814W} \equiv -2.5  ~ \log_{10}\ \frac{I_{\rm e^-}}{\rm 
  72} + 25.492 - 0.25 - 0.079$$ 
[Note:\  1)   $-2.5  ~  \log_{10}\   {I_{\rm  e^-}}$  is   simply  the
instrumental magnitude  (some codes, such  as DAOPHOT, add  a constant
value of 25 to the  instrumental magnitude to avoid negative numbers);
2) the adopted  normalization of the  fitting magnitude is  taken into
account by  the aperture correction $\Delta  m_{\rm PSF-AP({\it r})}$;
3) the sign of the last two quantities in each equation is negative.]
 
The calibrated CMDs are presented in Fig.\ 2. 
%
%
\begin{figure*} 
\includegraphics[width=15cm]{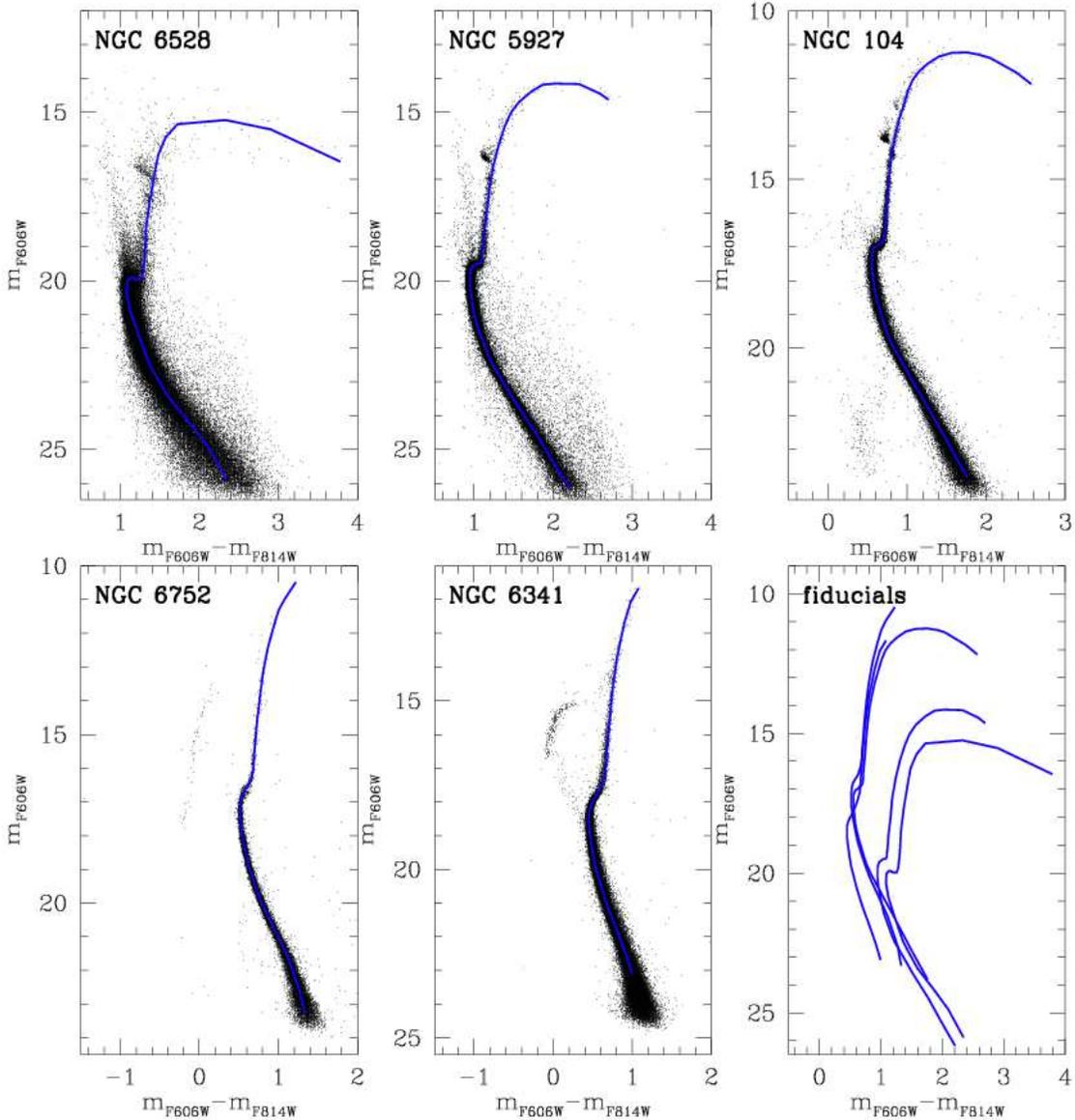} 
\caption{Observed CMDs for the five clusters discussed in the text and
summarized in Tab.\  6.  The fiducials are overplotted  onto each CMD,
and shown together in the bottom-right panel for comparison.}
\end{figure*} 
Superposed on the observed CMD are  the fiducial lines of the MS, SGB,
and RGB. The fiducial points which define them have been determined by
eye  and are given  in Table~7.   Note that,  due to  the differential
reddening in  the field, the fiducial  lines in the  cases of NGC~6528
and NGC~5927 are poorly defined.
 
These fiducial  points can be used  as a reference  to compare against 
those measured for other globular clusters or galaxies whose stars are 
resolved (Brown et al.\ 2003).

%
%
\begin{table}  
\begin{center}  
\begin{tabular}{c|cc}  
 Cluster  &  F606W data      &  F814W data        \\  
\hline 
 NGC 6528 &  4s,   50s, 450s & 1s,   20s,   350s  \\  
 NGC 5927 &  2s,   30s, 500s & 0.7s, 15s,   340s  \\  
 NGC  104 &  0.5s,  6s,  70s & 0.5s,  5.5s,  72s  \\  
 NGC 6752 &  0.5s,  4s,  40s & 0.5s,  4s,    45s  \\  
 NGC 6341 &  0.5s,  5s,  90s & 0.5s,  6s,   100s  \\  
\end{tabular}  
\end{center}  
\caption{Data set of the observed clusters. } 
\label{database}  
\end{table}   
%
%
%
%
\begin{table*}  
\begin{center}  
{\tiny 
%
\begin{tabular}{c}  
           NGC 6528 \\ 
\begin{tabular}{cc}  
  Colour & Mag      \\ 
\hline 
   3.79 &   16.46 \\  
   2.91 &   15.52 \\  
   2.33 &   15.24 \\  
   1.72 &   15.36 \\  
   1.57 &   15.75 \\  
   1.48 &   16.26 \\  
   1.43 &   16.85 \\  
   1.38 &   17.59 \\  
   1.33 &   18.41 \\  
   1.31 &   19.33 \\  
   1.28 &   19.82 \\  
   1.26 &   19.97 \\  
   1.20 &   19.97 \\  
   1.14 &   19.89 \\  
   1.09 &   20.02 \\  
   1.08 &   20.35 \\  
   1.12 &   20.86 \\  
   1.23 &   21.58 \\  
   1.36 &   22.45 \\  
   1.60 &   23.40 \\  
   1.85 &   24.14 \\  
   2.10 &   24.88 \\  
   2.35 &   25.88 \\  
\end{tabular}  
\end{tabular} 
\begin{tabular}{c}  
           NGC 5927 \\ 
\begin{tabular}{cc}  
  Colour & Mag      \\ 
\hline 
   2.70 &   14.63 \\  
   2.59 &   14.45 \\  
   2.33 &   14.17 \\  
   2.03 &   14.15 \\  
   1.87 &   14.18 \\  
   1.72 &   14.38 \\  
   1.57 &   14.69 \\  
   1.46 &   15.02 \\  
   1.39 &   15.40 \\  
   1.31 &   15.93 \\  
   1.24 &   16.59 \\  
   1.18 &   17.18 \\  
   1.15 &   17.84 \\  
   1.12 &   18.53 \\  
   1.12 &   18.94 \\  
   1.11 &   19.21 \\  
   1.10 &   19.32 \\  
   1.08 &   19.44 \\  
   1.04 &   19.49 \\  
   1.01 &   19.52 \\  
   0.97 &   19.64 \\  
   0.95 &   19.75 \\  
   0.95 &   20.00 \\  
   0.95 &   20.33 \\  
   0.98 &   20.66 \\  
   1.01 &   21.09 \\  
   1.08 &   21.63 \\  
   1.23 &   22.44 \\  
   1.51 &   23.58 \\  
   1.77 &   24.47 \\  
   1.99 &   25.36 \\  
   2.21 &   26.18 \\  
\end{tabular}  
\end{tabular} 
\begin{tabular}{c}  
           NGC 104 \\ 
\begin{tabular}{cc}  
  Colour & Mag      \\ 
\hline 
   2.57 &   12.17 \\  
   2.38 &   11.83 \\  
   2.05 &   11.39 \\  
   1.92 &   11.30 \\  
   1.75 &   11.23 \\  
   1.56 &   11.25 \\  
   1.40 &   11.36 \\  
   1.27 &   11.55 \\  
   1.16 &   11.78 \\  
   1.07 &   12.03 \\  
   1.01 &   12.35 \\  
   0.94 &   12.84 \\  
   0.88 &   13.32 \\  
   0.83 &   13.83 \\  
   0.80 &   14.26 \\  
   0.76 &   14.90 \\  
   0.74 &   15.55 \\  
   0.73 &   15.98 \\  
   0.71 &   16.37 \\  
   0.71 &   16.58 \\  
   0.70 &   16.74 \\  
   0.69 &   16.81 \\  
   0.67 &   16.90 \\  
   0.65 &   16.95 \\  
   0.62 &   16.99 \\  
   0.59 &   17.06 \\  
   0.57 &   17.16 \\  
   0.56 &   17.30 \\  
   0.56 &   17.52 \\  
   0.56 &   17.76 \\  
   0.57 &   18.01 \\  
   0.59 &   18.33 \\  
   0.62 &   18.67 \\  
   0.67 &   19.07 \\  
   0.73 &   19.50 \\  
   0.82 &   19.96 \\  
   1.02 &   20.77 \\  
   1.18 &   21.45 \\  
   1.47 &   22.65 \\  
   1.76 &   23.82 \\  
\end{tabular}  
\end{tabular} 
\begin{tabular}{c}  
           NGC 6752 \\ 
\begin{tabular}{cc}  
  Colour & Mag      \\ 
\hline 
   1.22 &   10.48 \\  
   1.08 &   10.99 \\  
   1.00 &   11.32 \\  
   0.94 &   11.80 \\  
   0.87 &   12.42 \\  
   0.81 &   13.19 \\  
   0.77 &   13.88 \\  
   0.73 &   14.60 \\  
   0.71 &   15.10 \\  
   0.69 &   15.60 \\  
   0.68 &   15.91 \\  
   0.66 &   16.15 \\  
   0.65 &   16.35 \\  
   0.64 &   16.44 \\  
   0.61 &   16.51 \\  
   0.58 &   16.66 \\  
   0.55 &   16.78 \\  
   0.53 &   16.87 \\  
   0.52 &   17.02 \\  
   0.51 &   17.16 \\  
   0.51 &   17.33 \\  
   0.52 &   17.61 \\  
   0.53 &   17.83 \\  
   0.56 &   18.26 \\  
   0.60 &   18.67 \\  
   0.66 &   19.17 \\  
   0.78 &   19.93 \\  
   0.93 &   20.63 \\  
   1.12 &   21.56 \\  
   1.27 &   22.61 \\  
   1.33 &   23.31 \\  
\end{tabular}  
\end{tabular} 
\begin{tabular}{c}  
           NGC 6341 \\ 
\begin{tabular}{cc}  
  Colour & Mag      \\ 
\hline 
   1.08 &   11.67 \\  
   0.98 &   12.06 \\  
   0.90 &   12.70 \\  
   0.80 &   13.73 \\  
   0.74 &   14.77 \\  
   0.72 &   15.41 \\  
   0.69 &   16.06 \\  
   0.68 &   16.54 \\  
   0.67 &   16.86 \\  
   0.66 &   17.07 \\  
   0.65 &   17.22 \\  
   0.63 &   17.34 \\  
   0.60 &   17.55 \\  
   0.57 &   17.68 \\  
   0.55 &   17.76 \\  
   0.52 &   17.91 \\  
   0.49 &   18.03 \\  
   0.48 &   18.12 \\  
   0.46 &   18.21 \\  
   0.45 &   18.30 \\  
   0.45 &   18.42 \\  
   0.45 &   18.58 \\  
   0.45 &   18.74 \\  
   0.46 &   18.93 \\  
   0.48 &   19.18 \\  
   0.49 &   19.43 \\  
   0.52 &   19.75 \\  
   0.55 &   20.08 \\  
   0.60 &   20.44 \\  
   0.66 &   20.90 \\  
   0.72 &   21.29 \\  
   0.81 &   21.85 \\  
   0.92 &   22.51 \\  
   1.00 &   23.11 \\  
\end{tabular}  
\end{tabular} 
} 
\end{center}  
\caption{Fiducial points in the plane:  
($m_{\rm F606W}-m_{\rm F814W}$), $m_{\rm F606W}$.  } 
\label{Fid}  
\end{table*}  
 
%
\subsection{Comparison with the Models}  

In this  section we compare the theoretical  models transformed to the 
ACS Vega-mag system with observed CMDs. 
 
Figure 3 displays  our adopted $\alpha$-enhanced  isochrones and ZAHBs 
compared   with  the observed  CMDs   of the  GGCs   47~Tuc (NGC~104), 
NGC~6752, and M92~(NGC~6341).  These clusters  span a wide metallicity 
range ($\rm -2.2 <  [Fe/H] < -0.6 $).   We selected these clusters for 
the comparison because    of their very  low  extinction  and lack  of 
differential reddening. 
 
In the following discussion, in  order to make the comparison with the 
values in  the literature more straightforward, we  give the reddening 
$E(B-V)$  and   the  apparent  distance   modulus  in  the   $V$  band 
($y_V$). The $E(B-V)$ comes from $E(m_{\rm F606W}-m_{\rm F814W}) \!=\! 
A_{\rm  F606W} -  A_{\rm F814W}$  as adopted  in the  best  fitting by 
interpolation of the values in Table 3.  The $y_V$ comes from the best 
fitting  distance  modulus  in   the  F606W  band  ($y_{\rm  F606W}$): 
$y_V\!=\!y_{\rm    F606W}   -    A_{\rm   F606W}    +    A_V$,   where 
$A_V\!=\!3.1E(B-V)$. 
 
In the case of  47~Tuc we used $\alpha$-enhanced ($<[\alpha/Fe]>=0.4$)
isochrones with  $\rm [Fe/H]\!=\!-0.66$  (Z$=$0.008), a value  in very
good agreement with spectroscopic measurements (see, e.g., Percival et
al.\ 2002,  Gratton et al.\ 2003);  the best match is  obtained for an
apparent distance  modulus $y_V\!=\!13.45$,  an age of  10 Gyr,  and a
reddening  $E(B-V)\!=\!0.04$ (the  8 and  13 Gyr  isochrones  are also
shown).
These values of distance   modulus and reddening are   consistent with 
empirical determinations by Percival   et al.\ (2002) and  Gratton  et 
al.\ (2003). 
 
The quality  of the fit is  satisfactory along most  of the isochrone,
worsening at the redder end of  both the RGB and MS.  The deviation of
the   lower    part   of   the   sequence   from    the   models   for
intermediate-metallicity  clusters is  well known  (see Bedin  et al.\
2001).
 
For  NGC~6752   we   have   employed    our  isochrones   with    $\rm 
[Fe/H]\!=\!-$1.26  (Z$=$0.002),  a   value close  to   the estimate by 
Rutledge et  al.\ (1997) with  their  calibration of the  CaII triplet 
index on  the  GGC Carretta \& Gratton  (1997)  metallicity scale.  We 
obtain  a best    match with $y_V$$=$13.25,   an age   of  12 Gyr  and 
$E(B-V)$$=$0.04 (the 10 and 14 Gyr isochrones are also shown). 
 
Again, the reddening agrees with independent empirical estimates (e.g. 
Gratton et  al.\ 2003), whereas the distance  is slightly  longer than 
the   MS-fitting  estimates (Carretta et  al.\   2000, Gratton et al.\ 
2003). 
These  empirical estimates  employ a  metallicity $\sim$0.2  dex lower 
than our  adopted value; it is  interesting to note,  judging from the 
numerical experiment performed by  Carretta et al.\ (2000, their Fig.\ 
5), that their empirical MS-fitting distance would come into agreement 
with our value if our  assumed metallicity is used in their MS-fitting 
technique. 
 
Had we used our isochrones for $\rm [Fe/H] \!=\!-$1.57 (Z$=$0.001) and
a value close  to the metallicity estimate by  Rutledge et al.\ (1997)
obtained  calibrating the  calcium index  on the  Zinn \&  West (1984)
metallicity scale, the quality of the fit to the RGB location would be
worse   (isochrone  too  blue),   the  age   older  by   $\sim$1  Gyr,
$E(B-V)\sim$0.02 mag higher and $y_V \sim$0.07 mag larger.
 
The comparison with  the M~92 (NGC~6341) CMD has  been performed using
isochrones  with [Fe/H]$\approx  -$2.1 (Z$=$0.0003),  a value  in fair
agreement with the estimate by Carretta \& Gratton~(1997) ($\rm [Fe/H]
\!=\!-2.16$).  The  best agreement is  achieved for an age  of 12~Gyr,
$E(B-V)$$=$0.04 and $y_V\!=\!14.85$ (the  10 and 14 Gyr isochrones are
also shown).
 
%
%
\begin{figure*} 
\includegraphics[width=15cm]{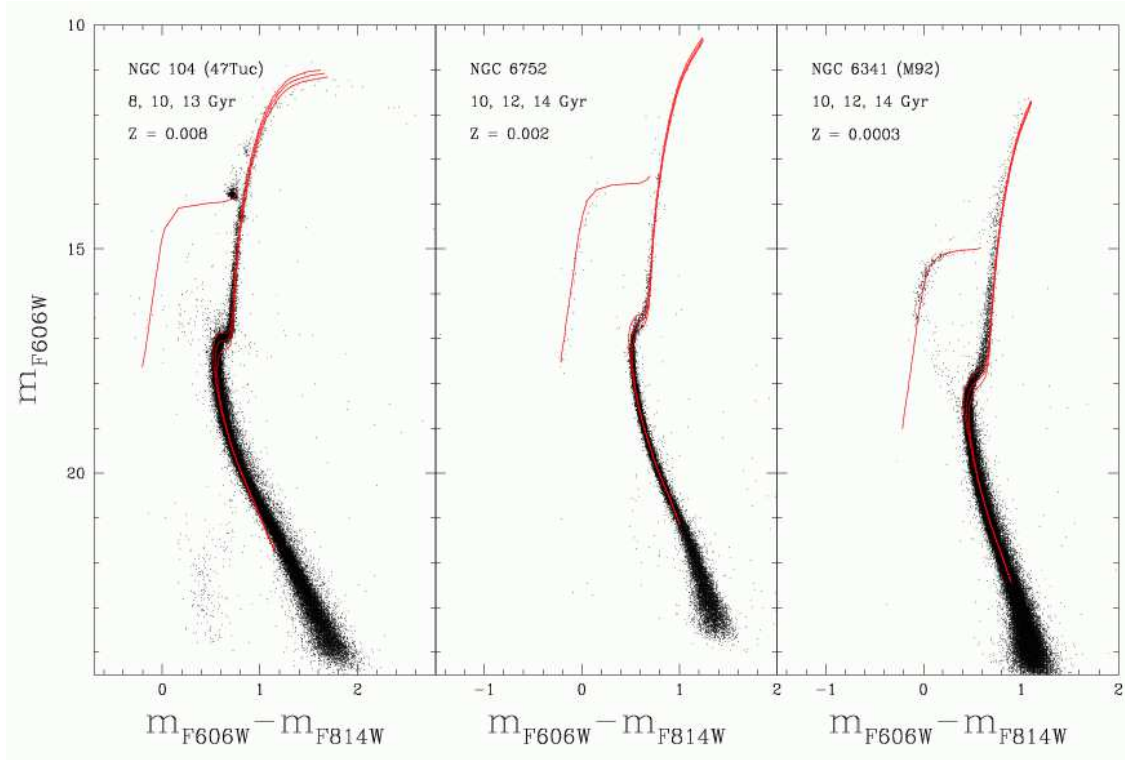}
\caption{Comparison  of three  GGCs CMDs  (M~92, NGC~6752  and 47~Tuc) 
with the corresponding isochrones (see text for details).} 
\end{figure*} 
 
%
\section{SUMMARY}  
%
%
In  this  paper  we  have  proposed a  simple  zero-point  photometric 
calibration of the ACS/WFC data, based on a model spectrum of Vega and 
the most up  to date in-flight transmission curves  of the ACS camera. 
We  have named  this new  in-flight system  the {\em  WFC/ACS Vega-mag 
photometric  system}.  The  proposed  calibration is  accurate at  the 
level of a  few hundredths of a magnitude.  We  give the basic recipes 
to calibrate both the observed data  and the models, and present a set 
of models by Pietrinferni et  al.\ (2004) already transformed into the 
new ACS Vega-mag system. 
 
Since the primary  purpose of this investigation is  to make available 
to  the  public  a  complete  set of  stellar  models  and  isochrones 
transformed  to  the  ACS  Vega-mag  system,  we  have  developed  the 
following web site where all  models and isochrones can be retrieved:\ 
{\sf http:// www.te.astro.it /BASTI /index.php}. 
 
We show CMDs for 5  Galactic globular clusters, spanning a metallicity 
range $-2.2<$[Fe/H]$<-0.04$;  these CMDs  have been obtained  from ACS 
observations, calibrated to the ACS  Vega-mag system, and used to test 
our models. 
We also  give  the  fiducial points  representing their RGB,  SGB, and 
Main  Sequence from  the red  giant  branch tip to  several magnitudes 
below the  TO. These CMDs  and fiducial points will  be useful for the 
study of  the old stellar populations in  other stellar systems (Brown 
et al.\ 2003). 
 
 
\section*{Acknowledgments} 
 
We  are very  thankful to  Marco Sirianni  for providing  us  with the
ACS/WFC transmission  curves, and also for discussions.   We thank the
anonymous referee  for useful suggestions, James Manner  for a careful
reading of the manuscript, and Phil James for polishing it.  This work
has been partially  supported by the Agenzia Spaziale  Italiana and by
the  Ministero  dell'Istruzione,  Universit\`a  e  Ricerca  under  the
program PRIN2003.

  

\begin{thebibliography}{} 
%
\bibitem []{} Alexander D.\ R., Ferguson J.\ W., 1994, ApJ, 437, 879  
\bibitem []{} Anderson J., King I.\ R., 2000, PASP, 112, 1360 
\bibitem []{} Angulo C., Arnould M., Rayet M., et al. (NACRE Collaboration), 
                1999, Nucl.\ Phys.\ A, 656, 3 
\bibitem []{} Bahcall J.\ N., Serenelli A.\ M., Pinsonneault M., 2004, ApJ,  
		{\sl submitted}, (astro-ph/0403604)  
\bibitem []{} Bedin L.\ R., Anderson J., King I.\ R., Piotto G., 2001 ApJ, 560, L75 
\bibitem []{} Bessell M.\ S., 1990, PASP, 102, 1181 
\bibitem []{} Bessell M.\ S., Castelli F., Plez B., 1998, A\&A, 333, 231 
\bibitem []{} B\"ohm-Vitense E., 1958, Z. Astrophys., 46, 108 
\bibitem []{} Brown. T. et al., 2003, ApJ, 592, L17 
\bibitem []{} Carretta E., Gratton R.\ G., 1997, A\&AS, 121, 95   
\bibitem []{} Cassisi S., Salaris M., 1997, MNRAS, 285, 593 
\bibitem []{} Cassisi S., Salaris M., Irwin A.\ W., 2003, ApJ, 588, 862 
\bibitem []{} Cassisi S., Castellani V., degl'Innocenti S., Weiss A., 1998, A\&AS, 129, 267 
\bibitem []{} Castellani V., Chieffi A., Pulone L., Tornamb\'e A.,   
		1985, ApJ, 296, 204  
\bibitem []{} Castelli F., Kurucz R.\ L., 1994, A\&A 281, 817 
\bibitem []{} Castelli F., Kurucz R.\ L., 2003, Proceed. of IAU Symp.\ 210, 
                Modelling of Stellar Atmospheres, eds.\ N.\ Piskunov et al., 
                poster A20 on the enclosed CD-ROM, (astro-ph/0405087) 
\bibitem []{} Chaboyer B., 1995, ApJL, 444, 9 
\bibitem []{} Cox J.\ P., Giuli R.\ T., 1968,  
                in \lq{Principles of Stellar Structure}\rq, Gordon \& 
		Breach, London, Vol.\ II 
\bibitem []{} De Marchi G., Sirianni M., Gilliland R., Bohlin R., Pavlosky C.,  
              Jee M., Mack J., van der Marel R. \& Boffi F. Instrument Science Report ACS 2004-08   
\bibitem []{} Graboske H.\ C., Dewitt H.\ E., Grossman A.\ S., 
		Cooper M.\ S., 1973, ApJ, 181, 457 
\bibitem []{} Gratton R.\ G., Bragaglia A., Carretta E., 
     	        Clementini G., Desidera S., Grundahl F., 
		Lucatello S., 2003. A\&A, 408, 529 
\bibitem []{} Grevesse N., Noels A., 1993, in Origin and Evolution of the 
                Elements, ed.\ N.\ Prantzos, E.\ Vangioni-Flam, M.\ Cass\'e  
		(Cambridge:\ Cambridge Univ.\ Press), 14 
\bibitem []{} Haft M., Raffelt G., Weiss A., 1994, ApJ, 425, 222 
\bibitem []{} Holtzman J.\ A., Burrows C.\ J., Casertano S.,  	 
		Hester J.\ J.,  Trauger J.\ T., Watson  A.\ M.,    
		Worthey G., 1995, PASP, 107, 1065 
\bibitem []{} Krishna-Swamy K.\ S., 1966, ApJ, 145, 174 
\bibitem []{} Krist J., Instrument Science Report ACS 2003-06   
\bibitem []{} Kunz R., Fey M., Jaeger M., Mayer A., Hammer J.\ W., Staudt G.,  
		Harissopulos S., Paradellis T.,  2002, ApJ, 567, 643 
\bibitem []{} Iglesias C.\ A., Rogers F.\ J., 1996, ApJ, 464, 943 
\bibitem []{} Landolt-B\"ornstein 1982, Numerical Data and Functional 
                Relationships in Science and Technology, vol.2, Astronomy and Astrophysics 
\bibitem []{} Mathis J. S., 1990, ARA\&A, 28, 37 
\bibitem []{} Momany Y., Cassisi  S.,  Piotto G.,  Bedin L.\  R., 
              Ortolani S.,  Castelli F.,  Recio-Blanco A.,  2003, 
              A\&A, 407, 303 
\bibitem []{} Percival S.\ M., Salaris M., van Wyk F.,  Kilkenny 
		D., 2002, ApJ, 573, 174 
\bibitem []{} Pietrinferni A., Cassisi S., Salaris M., Castelli 
		F., 2004, ApJ, in press, (astro-ph/0405193) 
\bibitem []{} Potekhin A.\ Y., 1999, A\&A, 351, 787 
\bibitem []{} Reimers D., 1975, Mem.\ Soc.\ R.\ Sci.\ L\'\i ege, 8, 369 
\bibitem []{} Riello M. et al. 2003, A\&A, 410, 553 
\bibitem []{} Riess A. \& Mack J. Instrument Science Report ACS 2004-006   
\bibitem []{} Rogers F.\ J., 2001, Contrib.\ Plasma.\ Phys., 41, 179 
\bibitem []{} Rutledge G.\ A., Hesser J.\ E., Stetson P.\ B. 1997, PASP, 109, 907 
\bibitem []{} Salaris, M., Riello M., Cassisi S., Piotto G., 2004, A\&A, 420, 911 
\bibitem []{} Sirianni M., De Marchi G., Gilliland R., Bohlin R., Pavlovsky C.,  
                Mack J., 2002, HST Calibration Workshop 2002,   
		ed.\ S.\ Arribas, A.\ Koekemoer, B.\ Whitmore  
		(Baltimore:\ STScI), p.\ 13  
\bibitem []{} Sirianni, M., Jee, M., Ford, H., Illingworth, G., Clampin, M., 
              Hartig, G., De Marchi, G., Gilliland, R., Benitez, N., Blakeslee, J., 
              Mack, J., Martel, A., and McCann W.J. 2004, submitted to PASP 
\bibitem []{} Spergel D.\ N., Verde L., Peiris H.\ V.\ et al., 2003, ApJS, 148, 175 
\bibitem []{} Zinn R., \& West M.\ J., 1984, ApJS, 55, 45 
%
\end{thebibliography}
\end{document}